# Effects of size of an oxide layer on the surface properties of lithium irradiated by deuterium plasma

P. Krstic
*TheoretiK, Port Jefferson Station, NY 11776, USA and*
*Department of Material Science and Chemical Engineering, Stony Brook University, Stony Brook, NY USA*

A. Maan, R. Majeski
*Princeton Plasma Physics Laboratory, Princeton, NJ, USA*

B.E. Koel
*Department of Chemical and Biological Engineering, Princeton University, Princeton, NJ, USA*
(Dated: September 13, 2024)

We investigate how growth of an oxide film will influence the deuterium recycling properties of Li on plasma facing surfaces. Lithium films on the walls or plasma-facing material surfaces of a fusion vacuum vessel improves plasma performance in part by removing residual impurity atoms from the plasma. Oxygen atoms, from residual water vapor or eroded oxide surfaces, are a prominent ubiquitous impurity that can create a layer of oxide on lithium films. The results of the work herein also indicate a possibility for determining the thickness of the oxide layer on such Li films by measuring the reflection probability of incident deuterium particles from the surface.

## INTRODUCTION

We aim to explore the molecular dynamics of incident deuterium (D) ions with varying impact energies perpendicular on a surface that is oxidized lithium of varying thickness, ranging up to a few nanometers, on top of elemental lithium. This paper documents how oxide growth on top of a lithium film will change the D retention or recycling capability of these plasma-facing component (PFC) coatings. In addition, the results provide a pos- sible new approach for an experimental determination of the oxide layer thickness at the film surface.

The interaction of hydrogenic ions with lithium-based surfaces is a topic of significant interest in the field of fusion energy research. The conditioning of PFCs with lithium has been shown to improve plasma performance in tokamak operations, in part because lithium is ex- pected to retain hydrogen and impurities from the plasma [1–4]. A key aspect of this conditioning process is the in- teraction between residual water vapor in the tokamak vacuum vessel and the lithium or lithium compound sur- faces. Specifically, lithium conditioning experiments on LTX and LTX-β have shown that the applied lithium starts to oxidize at a rate dictated by the residual vac- uum conditions of the vacuum vessel [5–7].

Oxidation of lithium PFCs is an inescapable phe- nomenon for any fusion reactor that chooses to employ lithium as a first wall. Oxidation starts with the cre- ation of oxide monolayers that grow in thickness with exposure to $H_2O$ and any O-containing residual gases in the vacuum vessel. This kind of growth can be mod- elled by a metal oxide growth kinetic model [8], which states that, in general, the oxide growth kinetics can be modeled by an advection-diffusion equation. At the on- set, oxide growth is driven by advection in the thin film limit [9]. A potential develops across the oxide film that drives the metal ions to the gas-oxide interface and oxy- gen ions to the metal-oxide interface. As the oxide grows in thickness, the potential across the film is shielded by free electrons, and eventually, when the oxide grows thick enough, further growth is dominated by Fickian diffusion in the thick film limit [10]. Using data from experiments on LTX-β [5–7], the $Li_2O$ growth rate for lithium-coated PFCs can be estimated to be on the order of 0.5 nm/hr. Lithium on the walls of a fusion reactor is expected to im- prove performance in two ways, i.e., by removing higher atomic number impurities present in the residual vac- uum and retaining incoming hydrogenic ions. Impurities sequestered by lithium leave behind a cleaner base vac- uum and therefore reduce radiative power loss from the plasma. Retaining incoming charged hydrogenic ions de- creases recycling and the population of cold hydrogenic species at the edge of the plasma that act as an energy sink and lowers the edge temperature primarily through charge exchange [11–15].

The results given herein emerge from, and uses expe- rience out of a series of our previous molecular dynamics studies on the physical response and chemistry of amor- phous surfaces of lithium [19, 20], lithium-hydride [16], and lithium oxide and hydroxide [17, 18] when irradiated by hydrogen ions from fusion plasma.

## COMPUTATIONAL METHOD

We created six layers of a composition pertaining to lithium-oxide starting from amorphous slabs of 20,000 - 30,000 Li atoms, and then accumulating oxygen atoms at the surface in a layer-by-layer manner. Each new layer of 100 randomly distributed O atoms was placed 0.5-

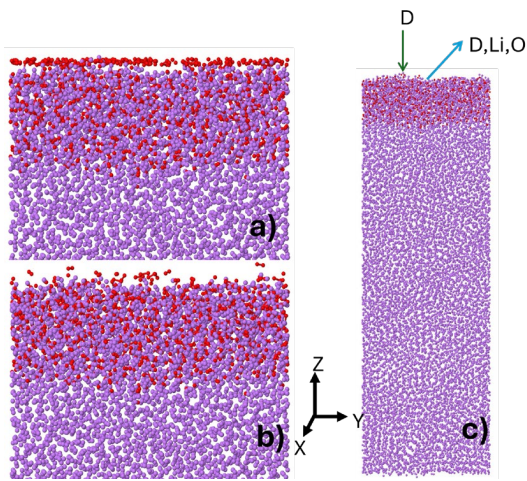

FIG. 1. Creation of the Li oxide layers. a) A thin layer of oxygen (red) atoms that were randomly distributed close to the Li (purple) surface. b) Energy minimization of the whole system was carried out. c) Thermalization of the entire system to 300 K was done and then relaxed for 50 ps. This surface was then irradiated by D atoms, which were either reflected or retained or caused sputtering of Li and O atoms.

1 Å above the top surface of dimensions 25x25 Å$^2$, as shown in Fig. 1a. The energy of the whole system was then minimized (Fig. 1b), followed by slow thermalization to 300 K using a Langevin thermostat and relaxation for 50 ps (Fig. 1c). This procedure was repeated until the Li-O layer reached the desired thickness. Using this approach, we created slabs of various oxide layer thickness, which together with reference cases of pure, bulk a:Li and a:Li$_2$O, were used to determine the reflection and sputtering probabilities during non-cumulative D irradiation of these samples for D incident along the direction of the Z-axis over a range of impact energies of 5-300 eV (Fig. 1c).

The creation of surface oxide samples 1-5 (via optimization, thermalization, and relaxation), as well as calculations of the reflection and sputtering probabilities in Section 3 was done by classical molecular dynamics (CMD) using the Large-scale Atomic/Molecular Massively Parallel Simulator (LAMMPS) and utilizing Reax Force Field (ReaxFF) for Li-O-H systems, which were developed and tested in our previous publications [16–20].

An important characteristic of a Li-O-H system is the large disparity of the electronegativities [21] of Li (0.98) and O (3.44), with H being in the middle (2.20). A consequence of this imbalance of electronegativities is that a strong mutual polarization of the atoms is involved (dependent on the instantaneous position of all atoms in the system) with the creation of polar-covalent interactions and bonds. Use of ReaxFF includes a possibility for dynamic polarizations and charging of atoms using the Electronegativity Equalization Method (EEM), making ReaxFF [22] well-suited for description of the Li-O-H atomic system dynamics [16–20].

One aspect of the present study is the significant extension of the highest D impact energy E in CMD to 300 eV. For each of the 8 impact energies used in the range of 5-300 eV, we impinged 1,000 D atoms on each sample and generated trajectories, which were shown to yield a satisfactory maximal standard error for the calculated reflection probabilities.

## RESULTS

The thickness of the oxide layers at the a:Li surface was determined by dividing the slab in the Z-direction into 2 Å sub-layers, each of them containing $n_O$ oxygen atoms and $n_{Li}$ lithium atoms. The zero of the Z-axis was set at the top surface layer of Li. We determined distribution of the ratio $n_O/n_{Li}$ at the midpoint of the corresponding sub-layer for each oxide layer. Each distribution was normalized to unity at its highest value, as shown in the inset of Fig. 2. The thickness of an oxide layer was defined as the depth of a $n_O/n_{Li}$ ratio at 0.5 of its highest value. Thus, sample 1 had an oxide thickness of 0.74 nm, while the thickest oxide layer was 2.83 nm for sample 5. The oxide thickness values, $t_R$, for five different samples are shown in the table inset in Fig. 2. As found in our previous publications [16, 17, 19, 20], the D reflection depth, $d_R$, in bulk a:Li is much larger than that in bulk a:Li$_2$O. Oxidation to form thin oxide layers at the surface results in $d_R$ values that fall between those for bulk a:Li and bulk a:Li$_2$O.

The reflection probabilities in Fig. 2 for all 5 samples are between the corresponding values for a:Li and a:Li$_2$O, as expected. Moreover, the reflection probabilities follow the order defined by the oxide thickness, from smaller to larger. Since distributions of the oxygen atoms in the 5 samples are not identical, the reflection probability $P_R$ values in Fig. 2 indicate that the details of the oxide structure are not a decisive factor in determining the reflection probability, but rather is the thickness of the surface oxide on lithium.

The shapes of the curves in Fig. 2 can be better understood by analyzing the normalized reflection depth distributions at various values of oxide thickness and D impact energies, as shown in Fig. 3. Their cumulative counts normalized to 100 % were used to define $d_R$'s for which 95 % of reflection depths are smaller, for various values of $t_R$ and E, and these are presented in Fig. 3 as curves $d_R(E)$, parameterized by $t_R$'s. Since oxygen, being heavier, is a better "reflector" of impacting D than Li, the reflection depth decreases as the oxygen layers become thicker. At smaller values of $t_R$ as in samples 1 and 2, the $d_R(E)$ curves mainly follow the slope of the a:Li curve, except at lower energies when these follow the shape of a:Li$_2$O curve. The $d_R(E)$ curves for the thicker oxide samples conform the a:$Li_2O$ curve at higher energies. All curves have a slightly reduced slope at the highest energies considered.

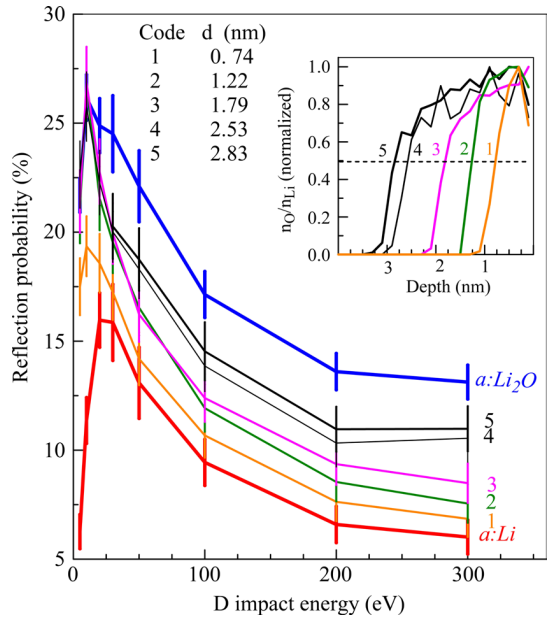

FIG. 2. Reflection probabilities of D atoms as a function of impact energy of D atoms impinging on surfaces denoted as 1-5 (thin lines), with an oxide thickness as given in the inset. Corresponding reflection probabilities from surfaces of pure Li and $Li_2O$ samples are shown by thick lines. Error bars present maximal standard errors derived from incident trajectory statistics.

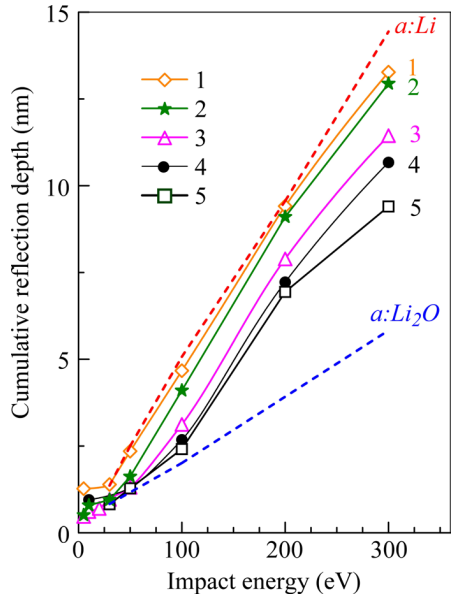

FIG. 3. Cumulative normalized values for the reflection depth at the level of 95 % for various impact energies of D and for several values of the oxide layer thickness. Dashed lines are results for pure bulk a:Li (red) and a:$Li_2O$(blue).

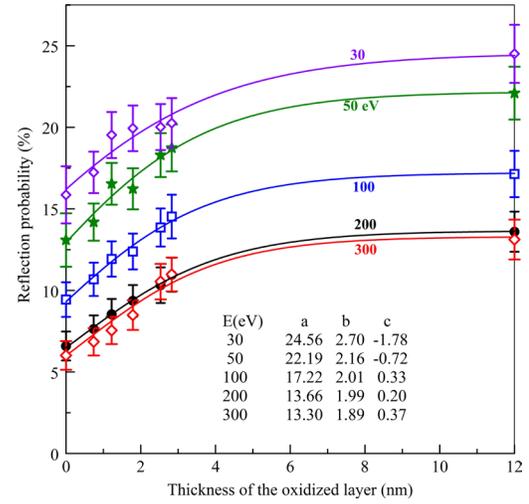

FIG. 4. D reflection probabilities for several Li oxide layer thickness values. The lines shown are curves fit to a sigmoidal function as defined in Eq. (1).

Fig. 4 shows the calculated reflection probabilities for several values of the thickness of the oxide layer, $t_R$, for various impact energies of incident D. The lines shown are analytical fits to the calculated data, and were obtained with a high correlation of 0.99 when using a 3-parameter sigmoidal type curves of the type,

$$P_R = \frac{a}{1 + \exp\frac{-(E-c)}{b}} \quad (1)$$

where E is in eV. Obviously, deuterium recycling capabilities, defined by its reflection and retention, for Li films will be strongly influenced by the formation of an oxide layer. The values of reflection probabilities at $t_R$=0 nm correspond to pure Li, while large $t_R$ values (12 nm in Fig. 4) correspond to $Li_2O$ surface layers. The $P_R$ values in the D impact energy range considered herein could change by factors of 1.5-2.5 due to oxidation of a Li film.

The parameters a, b and c in Eq. (1) are listed in Fig. 4. These curves, in addition, allow for estimation of a solid oxide layer thickness from measured data for the reflection probability of D at a fixed known impact energy, for a surface at 300 K. Since the parameters in Fig. 4 are smooth functions of energy, this enables obtaining information on the oxide thickness at any available D energy in the range of 30-300 eV. Thus, measuring the gas phase outcome of D reflection at a surface provides information on the oxide thickness at the solid surface, directly connecting two phases constituting the plasma-material interface.

This material is based upon work by the U.S. Department of Energy, Office of Science/Fusion Energy Sciences under Award Number DE-SC0019308 to Princeton University and through contract number DE-AC02-09CH11466 to the Princeton Plasma Physics Laboratory (PPPL). PK is grateful to Princeton University for using the Stellar HPC cluster, and to ACCESS of NSF for use of the SDSC Expanse HPC through grant TG-DMR110037. AM is grateful for many discussions with Robert Kaita about applications of lithium on the walls of fusion experiments.

---